\documentclass[11pt]{article}
\usepackage{moriond,epsfig,multirow,url}

\def\ra{\rightarrow}

\def\be{\begin{equation}}
\def\ee{\end{equation}}
\def\bea{\begin{eqnarray}}
\def\eea{\end{eqnarray}}

\begin{document}
\newcommand{\nc}{\newcommand}
\nc{\beq}{\begin{equation}}
\nc{\eeq}{\end{equation}}
\nc{\tb}{\rm tan \beta}
\nc{\tautau}{\rm H/A/h \ra \tau \tau}
\vspace*{4cm}
\title{Possibility of $\rm tan\beta$ measurement in CMS at LHC}

\author{ M. HASHEMI }

\address{Institute for Studies in Theoretical Physics and Mathematics(IPM),\\
Shahid Lavasani st., P.O.Box 19395-5531, Tehran, Iran}

\maketitle\abstracts{
The achievable $\rm tan\beta$ determination accuracy for CMS at LHC is presented. 
Using MSSM $\rm H/A \rightarrow \tau\tau$ decay in the associated production process
 $\rm gg \rightarrow b\bar{b}H/A$, the event rates are measured at large $\rm tan\beta$ and the systematic and statistical errors are estimated.
Due to sensitivity of the above event rates to $\rm tan\beta$, it is shown that it is possible to determine constraints on $\rm tan\beta$ for a given set of SUSY parameters and uncetainties.}

\section{Introduction}
Supersymmetry(SUSY)- a symmetry under interchange of bosonic and fermionic degrees of freedom- provides an elegant solution to the hierarchy problem, and thus has been considered as one of the most promising new physics scenarios among various possibilities. The Minimal Supersymmetric extention to Standard Model(MSSM) requires the introduction of two Higgs doublets in order to preserve supersymmetry. There are five physical Higgs particles, two CP-even (h,H), one CP-odd (A) and two charged ones ($\rm H^{\pm}$). All couplings and masses of the MSSM Higgs sector are determined at lowest order by two independent parameters, which are generally chosen as $\rm tan\beta = \upsilon_2 / \upsilon_1$, the ratio of the vacuum expectation values of the two Higgs doublets, and the pseudoscalar Higgs boson mass $\rm m_A$.
One of the most important parameters to be determined in MSSM is $\rm tan\beta$ since it enters in all sectors of the theory.
In this report possibility of determining $\rm tan\beta$ value is presented taking into account different systematic and statistic uncertainties of measurements \cite{Sami}. The results are presented for $\rm m_h^{max}$ benchmark scenario \cite{mhmax} with the following set of MSSM parameters:  
SU(2) gaugino mass $\rm M_2=200 GeV/c^2$, $\rm \mu=300 GeV/c^2$, gluino mass $\rm M_{\tilde{g}}=800 GeV/c^2$, SUSY breaking mass parameter $\rm M_{SUSY}= 1TeV/c^2$ and stop mixing parameter $\rm X_t=\sqrt{6}(X_t=A_t-\mu cot\beta)$. The top quark mass is set to $\rm 175GeV/c^2$. The Higgs boson decay to SUSY particles are allowed.\\   
\section{Summery of $\rm H/A \rightarrow \tau \tau$ analysis results and expected $5\sigma$ discovery reach}
\subsection{NLO cross section}
\begin{figure}[hbtp]
  \begin{center}
    \resizebox{8cm}{7cm}{\includegraphics{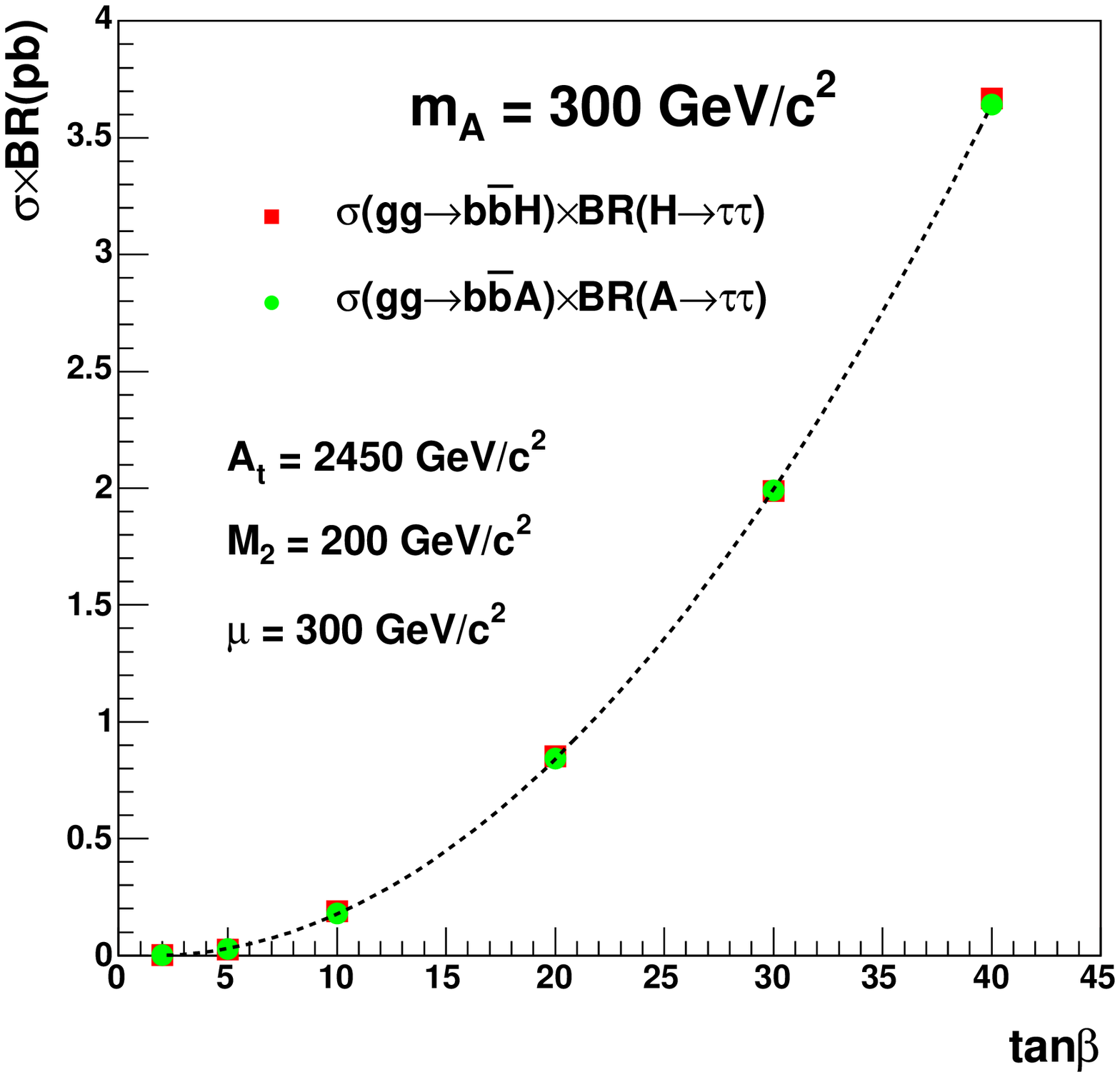}}
    \caption{cross section times branching ratio vs $\rm tan \beta $}
    \label{fig:ex1}
  \end{center}
\end{figure}

The NLO cross section calculation results show that at large $\rm tan \beta$ the cross section for $\rm gg\rightarrow b\bar{b}H/A/h$ exceeds that of $\rm gg \rightarrow H/A/h$ and Higgs bosons are produced predominantly in association with two b quarks \cite{nlo}.
Thus only $\rm gg \rightarrow b\bar{b}H/A/h$ has been considered as the production process.
Figure \ref{fig:ex1} shows NLO cross section times branching ratio versus $\rm tan \beta $.
Dominant parts of the cross section are proportional to $\rm tan^2 \beta $ at leading order. At NLO there are linear terms which can be combined together with leading order terms in a term as $\rm {tan^2 \beta}_{eff}$ which is denoted as $\rm tan \beta $ hereafter.
As a conclusion the uncertainity on the $\rm tan \beta $ measurement is half of uncertainity of rate measurement.     

\subsection{Signal and background identification and simulation tools}
Events are generated by PYTHIA6 \cite{pythia}, TopReX \cite{toprex}, TAUOLA \cite{tauola}, signal cross section is calculated using PPHTT \cite{pphtt} and Higgs branching ratios with HDECAY \cite{hdecay}. 
Four final states have been studied with their different braching fractions as listed in Table \ref{fractions}.
\begin{table}[htb]
\caption{Final states of $\rm H/A/h \ra \tau \tau $ with branching fractions, ($\rm \ell \ell = e \mu , \mu \mu , ee $)}
\label{fractions}
\begin{center}
\begin{tabular}{|l|c|}
  \hline
Final state & Branching ratio \\
\hline
$\rm H/A/h \rightarrow \tau \tau \rightarrow e\mu +X$ & $\rm \sim6.3\%$\\
\hline
$\rm H/A/h \rightarrow \tau \tau \rightarrow \ell \ell +X$ & $\rm \sim12.5\%$\\
\hline
$\rm H/A/h \rightarrow \tau \tau \rightarrow \ell j+X$ & $\rm \sim45.6\%$\\
\hline
$\rm H/A/h \rightarrow \tau \tau \rightarrow jj +X$ & $\rm \sim41.5\%$\\
\hline
\end{tabular}
\end{center}
\end{table}
Main common background events for all channels are :
$\rm Z,\gamma^* \rightarrow \tau \tau$ Drell-Yan process, $\rm t\bar{t} $ production with real and fake $\rm \tau$'s and single top production Wt. Channels with leptonic final states suffer from $\rm b \bar{b}$ events. W+jet process is the background for final states with hadronic $\rm \tau $ decays. When both $\rm \tau$'s decay hadronically there is an aditional QCD background with possibility of having fake $\rm \tau$'s.



\subsection{$\rm 5 \sigma$ discovery contours}
\begin{figure}[hbtp]
  \begin{center}
    \resizebox{10cm}{10cm}{\includegraphics{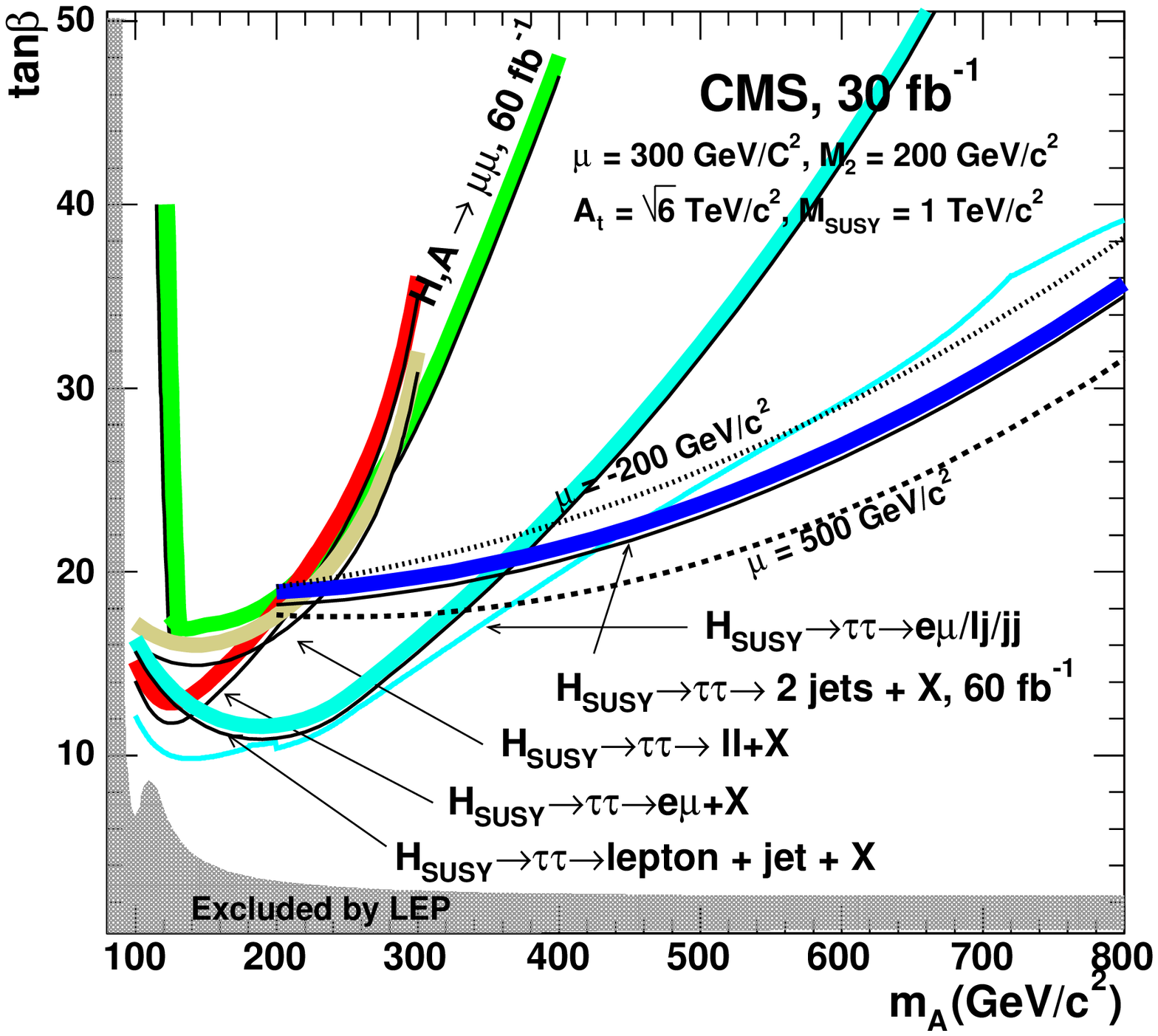}}
    \caption{$\rm 5 \sigma$ contour for different final states of $\rm H_{SUSY} \ra \tau \tau$.}
    \label{fig:fivesigma}
  \end{center}
\end{figure}

The $\rm 5 \sigma$ discovery contours for $\rm H/A/h \rightarrow \tau \tau $ with different final states of $\rm e\mu , \ell \ell $ and $\rm \ell j$ are shown in Figure \ref{fig:fivesigma} for integrated luminosity of $\rm 30fb^{-1}$. For two jet final state the $\rm 5 \sigma $ contour for $\rm 60fb^{-1}$ is shown. Also shown is the $\rm 5 \sigma $ discovery contour for $\rm H/A/h \rightarrow \mu \mu $ for $\rm 60 fb^{-1}$ . 
As is seen in the figure this channel is one of the most promising channels for heavy neutral MSSM Higgs boson discovery. 
\section{Statistical and systematic uncertainties of the production cross section measurement}
Systematic uncertainties of the measured production cross section come from the luminosity uncertainty, experimental selection uncertainty and background uncertainty.
The total uncertainty is the quadratic sum of statistical and systematic errors:
\begin{equation}  
\rm \frac{\Delta \sigma_{prod.}}{\sigma_{prod.}}= \frac{\sqrt{N_S+N_B}}{N_S}\oplus \frac{\Delta L}{L}\oplus \frac{\Delta \varepsilon_{sel.}}{\varepsilon_{sel.}} \oplus \frac{\Delta N_B}{N_S}
\end{equation}
where the first term is the statistical uncertainty and other terms are systematic errors mentioned above.
\subsection{Statistical and luminosity uncertainty}
Statistical uncertainty of the signal events is calculated by weighted summing over all final states.
Depending on $\tb$ and final state the statistical errors are 5-25$\%$.
The uncertainty of the luminosity measurement is assumed to be $\rm 5\%$
\subsection{Signal selection uncertainty}
The uncertainty of the signal selection efficiency comes from the calorimeter energy scale(since jets and missing $\rm E_t$ thresholds are required), b-tagging efficiency and $\rm \tau$-tagging efficiency.
The total selection efficiency uncertainty has been obtained to be $\rm \sim 4.3\%$.
\subsection{Background uncertainty}
The background contribution to the signal selection efficiency is estimated by fitting the invariant mass of two $\rm \tau$'s, $\rm m_{\tau \tau}$, and varying the number of signal and background events.
The error of the fit gives the background uncertainty.
The background uncertainty is estimated to be $\rm \Delta N_B/N_S=10\%$. So the total systematic uncertainty of production cross section is $\rm \sim 12 \%$ which is comparable with statistical uncertainty.

\section{Estimating $\rm tan \beta$ uncertainty}
As was mentioned earlier the signal production cross section is proportional to the square of $\rm tan \beta$ :
\begin{equation}
\rm \sigma_{prod.}=tan^2\beta \times X
 \end{equation}
so the error on $\rm tan \beta$ is :
\begin{equation}
\rm \frac{\Delta tan\beta}{tan\beta}=\frac{1}{2}(\frac{\Delta \sigma_{prod.}}{\sigma_{prod.}})\oplus \frac{1}{2}(\frac{\Delta X}{X})
\end{equation}
where $\rm \Delta X$ consists of theoretical uncertainty of the production cross section and branching ratios and cross section uncertainty due to the uncertainty of the Higgs boson mass measurement.

\subsection{Theoretical uncertainty of the signal selection}
According to NLO cross section calculations \cite{nlo}, the NLO cross section uncertainty for the signal is assumed to be $\rm 20-30\%$ for the total rate.
The branching ratio uncertainty is $\rm \sim 3\%$ which is due to SM parameters uncertainties.
\subsection{Uncertainty of the signal Higgs mass measurement}
The signal production cross section depends on the Higgs mass which is measured with some accuracy.
This induces some error on the cross section.
At $\rm 5 \sigma$ limit where the signal statistics is lowest, the mass measurement uncertainty brings $\rm 5-6\%$ uncertainty on $\rm tan \beta $ measurement.
\subsection{SUSY parameters uncertainy effects} 
SUSY parameters uncertainties are still unknown but to give an estimation of the rate sensitivity to SUSY parameters, those were varied by $\rm 20\%$ around the nominal values.
The variation of the rate within the discovery region is about $\rm 11\%$ which leads to at most $\rm 6\%$ uncertainty on $\tb$ measurement.
\section{$\tb$ measurement uncertainty in different final states of $\rm H/A \ra \tau \tau $}
\begin{figure}[h]
\centering
\vskip 0.1 in
\begin{tabular}{cc}
\begin{minipage}{7.5cm}
\centering
  \resizebox{\linewidth}{\linewidth}{\includegraphics{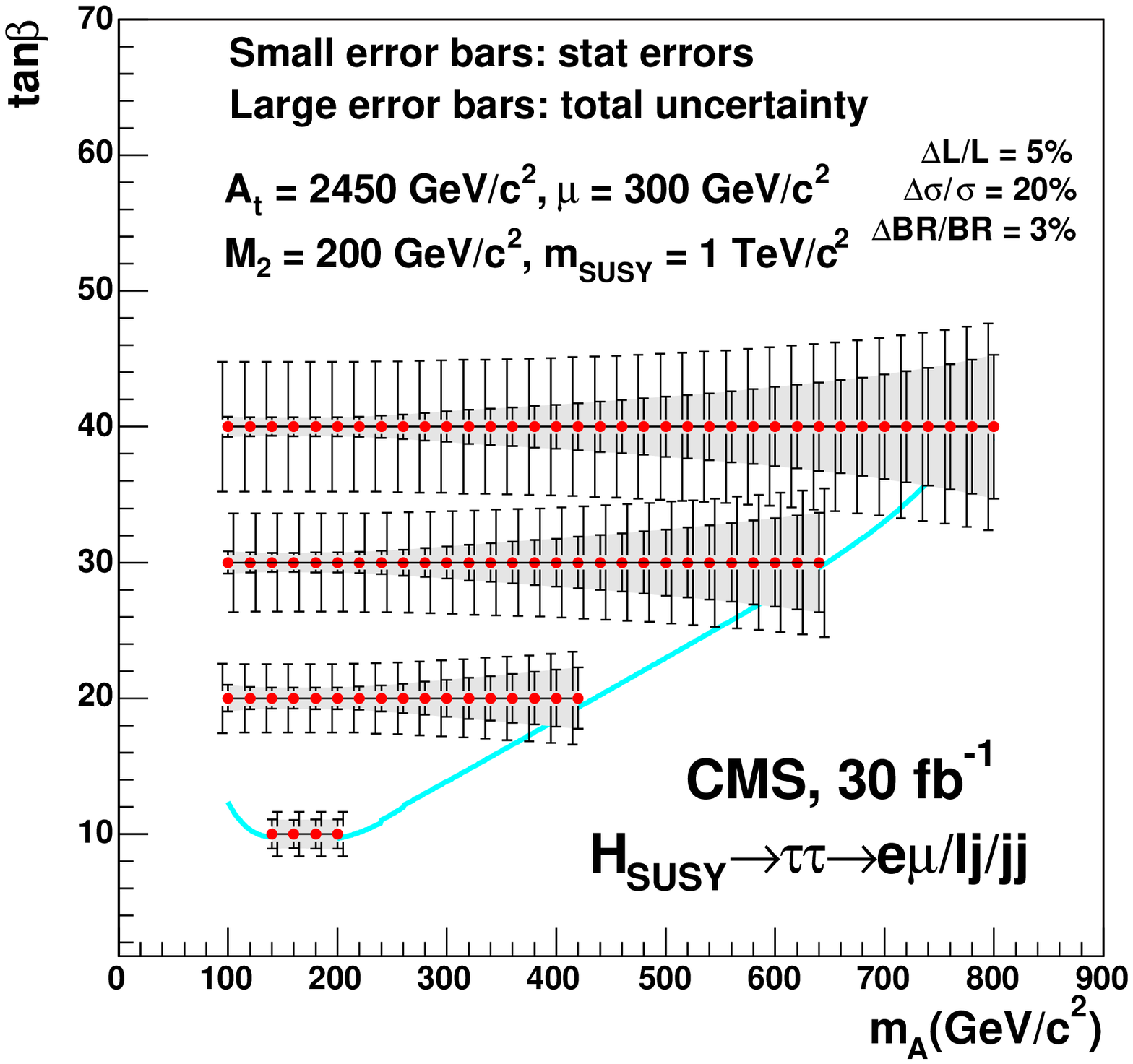}} 
  \caption{The uncertainty of $\tb$ measurement for $\rm 30fb^{-1}$}
  \label{fig:ex33} 
  \end{minipage}
&
\begin{minipage}{7.5cm}
\centering
  \resizebox{\linewidth}{\linewidth}{\includegraphics{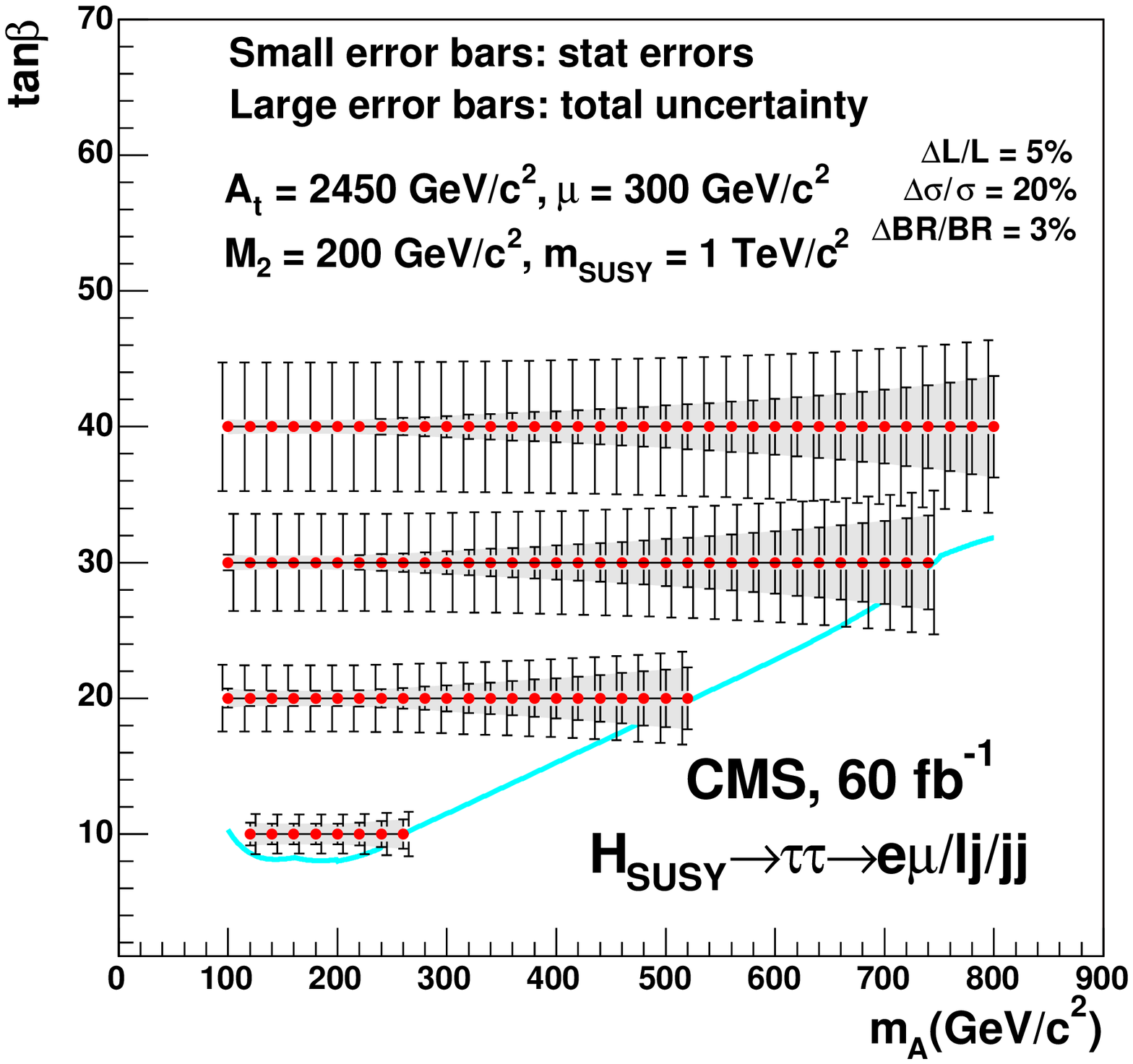}} 

\caption{The uncertainty of $\tb$ measurement for $\rm 60fb^{-1}$}
  \label{fig:ex44} 
\end{minipage}
\end{tabular}
\end{figure}

Figures \ref{fig:ex33},\ref{fig:ex44} show the total uncertainty as well as statistical uncertainties of $\tb$ measurement for $\rm 30fb^{-1}$ and $\rm 60fb^{-1}$ respectively.
    
\section{Conclusion}
The possibility of $\tb$ measurement in CMS at LHC was presented by estimating the precision of the cross section times branching ratio measurement in $\rm H/A \ra \tau \tau$ decay channel with different final states for $\rm 30fb^{-1}$.
The statistical uncertainty of production cross section is estimated to be $\rm 4-25 \%$ while the systematic error is $\rm \le 12\%$ which both depend on the signal significance.
Due to existence of large radiative corrections to the bottom Yukawa coupling, results presented correspond to $\tb_{eff}$ which absorbs the leading part of these corrections.
Close to the $\rm 5 \sigma $-discovery limit the statistical uncertainty is in the same order as the theoretical uncertainties but for $\tb$ regions where the signal significance is more than $\rm 5 \sigma$ significantly the theoretical errors dominate in the estimated total uncertainty of $\tb$ measurement.

\end{document}